\documentclass[aps,floatfix,twocolumn,superscriptaddress]{revtex4}
\usepackage{amsbsy}
\usepackage{amsmath}
\usepackage{epsfig}

\newcommand{\ket}[1]{|#1 \rangle}

\newcommand{\ketbra}[2]{\vert #1 \rangle \! \langle #2 \vert}

\newcommand{\be}{\begin{equation}}
\newcommand{\ee}{\end{equation}}
\newcommand{\bae}{\begin{eqnarray}}
\newcommand{\eae}{\end{eqnarray}}

\begin{document}

\title{Quantum chaos and operator fidelity metric}
\author{Paolo Giorda}
\affiliation{ Institute for Scientific Interchange, Viale Settimio Severo 65, I-10133 Torino, Italy}
\author{Paolo Zanardi}
\affiliation{Department of Physics and Astronomy,  Center for Quantum Information Science \& Technology, University of Southern California, Los Angeles, CA 90089-0484}
\affiliation{ Institute for Scientific Interchange, Viale Settimio Severo 65, I-10133 Torino, Italy}

\begin{abstract}
We show that the recently introduced operator fidelity metric provides a natural tool to investigate the cross-over to quantum chaotic behaviour. This metric is an information-theoretic measure of  the global stability of a unitary evolution against perturbations. We use random matrix theory arguments to conjecture that the  operator fidelity metric can be used as an "order parameter" to discriminates phases with regular behaviour from quantum chaotic ones. A numerical study of the onset of chaotic  in the Dicke model is given in order to support  the conjecture

\end{abstract}


\maketitle
{\em{ Introduction}}-- The information-theoretic task of distinguishing quantum states can be formulated in terms of the differential-geometric notion of {\em metric} over the quantum state space both in the pure and mixed state case \cite{woo-bra-ca}. This remarkable fact lies at the root of the recently proposed {\em fidelity approach} to quantum phase transitions (QPTs)
\cite{paukozanardi}.
The intuition behind this approach is quite simple: at the transition points a slight change of the system driving parameters e.g., external fields, gives rise to  a major
structural change of the associated (thermal or ground) quantum state and this, in turn, should result in an enhanced statistical distinguishability. In view of the above mentioned connection one expects detectable signature in the metric function e.g., singularities in the thermodynamical limit.
This intuition has been confirmed by the study of a host of different many-body systems, at zero  and finite temperature \cite{many-body,finite-T}.

In Refs \cite{opfid} the ground state approach has been extended to the operator level. In order to do that one has simply to notice that finite-time quantum evolutions over the Hilbert space $\cal H$ correspond to unitary operators that  belong to a space i.e., the space of linear operators ${\cal L}({\cal H})$, that can be turned itself into a Hilbert one in many different ways. This allows one to introduce the notion of {\em operator fidelity} and the associated metric  \cite{opfid}.

The main goal of this paper is to attempt to use operator fidelity  to investigate another fascinating and somewhat elusive  phenomenon : quantum chaos \cite{Chaos}. In many years of theoretical efforts a variety of  approaches, like random matrix theory \cite{RandomMtx}, quantum motion reversibility \cite{QCh-rev} and stability \cite{QCh-stab} up to very recent measures  of phase-space growth rate have been proposed \cite{casati08}. All of these related approaches  appear to be partially successful, yet a unified fully and  satisfactory characterization of quantum complex behaviour is still missing.

 In the following we first review the relevant background on operator fidelity metric and we then carry out our analysis along two well separated though  mutually supporting paths. In the first analytical part we explore the consequences of the physical intuition that is behind the approach that we advocate: the quantum chaotic evolutions can be characterized as those who have a resilience against random perturbations. This resilience is here quantified by the information-theoretic distance between perturbed and unperturbed evolutions. We will make use of random matrix theory \cite{RandomMtx} arguments to formulate and support our main conjecture on  operator metric (in)sensitivity of (chaotic) regular system to random perturbations.
 The second part of our paper  offers  an implementation these ideas in a particular system. More precisely, we numerically  analyze the cross-over to quantum chaos in  the Dicke model \cite{Dicke, DickeChaos} and we show how the operator metric approach can be  used as a tool to identify and characterize this cross-over. We will conclude by a summary and an outlook.

{\em{ Information-Theoretic metric over manifolds of unitaries}}-- Let us start by providing the setup and the basic facts about operator fidelity metric found in \cite{opfid}. Let $\cal H$ be a quantum state-space and $\rho$ a density matrix over it  ($\rho\ge 0,\, \rm{tr}\rho=1$).
 One can define the following hermitean product over ${\cal L}({\cal H})$:
$\langle X, Y\rangle_\rho:=\rm{tr}( \rho X^\dagger Y)
.$ If $\rho>0$ then this scalar product is non-degenerate and $\|X\|_\rho:=\sqrt{\langle X, X\rangle_\rho}$ defines a norm over ${\cal L}({\cal H}).$
The {\em operator fidelity} $X,Y\in{\cal L}({\cal H})$ is then given by
\begin{equation}
{\cal F}_\rho (X,Y):= |\langle X, Y\rangle_\rho|
\label{Eq.: op fid}
\end{equation}
This quantity has  well-defined operational meaning in terms of statistical distinguishability (the higher $\cal F$ the lower this latter) in terms of bi-partite quantum states \cite{distin}, interferometric schemes associated with $X,Y$ and $\rho$  and decoherence rates \cite{opfid}.
The case we will be  concerned with is when
$X=U_\lambda:=e^{-it H_\lambda},\,Y=U_{\lambda^\prime} (\lambda\neq\lambda^\prime)$  we will also  assume that $[\rho, H_{\lambda}]=0.$

In passing we would like to note that  (\ref{Eq.: op fid}) has relevance to non-equilibrium dynamics e.g., quenches where one goes suddenly from  $H_{\lambda}$ to
  $H_{\lambda}^\prime.$ Indeed  (\ref{Eq.: op fid}) is the modulus of
characteristic function of the work distribution function \cite{kur}
${\cal F}(U_\lambda(t),U_{\lambda^\prime}(t))=|\int d\omega  P(\omega) e^{i\omega t}|
$ where $P(\omega):=\sum_{m,n} \rho_{n,n} |\langle n|m^\prime\rangle|^2 \delta(\omega -E_n+E_{m^\prime}^\prime)$
($E_n$ and $|n\rangle$ and the corresponding primed quantities denote eigenvalues and eigenvectors of
$ H_{\lambda^\prime}$). Moreover for $\rho_{n,n}=\delta_{n,0}$ one obtains a  Loschmidt echo $|\langle 0|e^{-it H_{\lambda^\prime}}|0\rangle|$
pointing out to relevance of (\ref{Eq.: op fid}) to dephasing experiments \cite{losch, opfid}.

In this paper, following the differential-geometric spirit of \cite{paukozanardi} we are going to consider the operator fidelity (\ref{Eq.: op fid}) between infinitesimally different unitaries i.e., $\lambda^\prime=\lambda+ d\lambda.$
The proof of the following proposition is just a direct calculation analog to the one
performed at the state-space level \cite{paukozanardi}.
Let $\{ U_\lambda\}\subset {\cal U}({\cal H})$ be a smooth family of unitaries over $\cal H$
parametrized by elements $\lambda$ of a manifold $\cal M$. One finds ${\cal F}_\rho(U_\lambda, U_{\lambda+\delta\lambda})=1 -\delta\lambda^2/2
\chi_\rho(\lambda)(U^\prime,U^\prime )$
where $U^\prime=\partial U/\partial\lambda$ and
\begin{equation}
\chi_\rho(\lambda)(X,Y):=\langle X, Y\rangle_\rho -\langle X, U_\lambda \rangle_\rho\langle U_\lambda, Y\rangle_\rho
\label{G}
\end{equation}
We see that the leading term in the expansion of (\ref{Eq.: op fid})  defines a quadratic form
over the tangent space of the projective  ${\cal L}({\cal H})$ at $U_\lambda.$
For full rank $\rho$ that quadratic form is  a {\em metric}. 
This operator metric
(also referred to as operator fidelity susceptibility in \cite{opfid}) is the central tool of this paper.
The intuitive  meaning of $\chi$ is quite simple: the larger the operator metric at $U_\lambda$ the greater
the degree of statistical distinguishability between the quantum evolutions associated with two slightly different
set of control parameters $\lambda$ and $\lambda+\delta\lambda.$ This fact has been used in (\cite{opfid})
to use (\ref{G}) to study quantum criticality.

For the developments of this paper it is convenient to make use of the  {\em superoperator} formalism
common in Quantum Information Science.
 We define over  ${\cal L}({\cal H})$ the superoperator ${ L}_H:=[H,\bullet].$ This is nothing but  the generator of the Heisenberg evolution and it is easy to see that, if
$|n\rangle$'s and the $E_n$'s denote the eigenvectors and eigenvalues of $H(\lambda),$
$E_n-E_m$ and $|n\rangle\langle m|$ are the  eigenvalues and  eigenoperators of $L_H$ respectively.
 It follows that
$|\hat{\Psi}_{n,m}\rangle=\rho_{m,m}^{-1/2} |n\rangle\langle m|$ it is an {\em orthonormal} basis of ${\cal L}({\cal H})$
i.e., $\langle\hat{\Psi}_{n,m}|\hat{\Psi}_{p,q}\rangle=\delta_{n,p}\delta_{m,q}.$
If we  define $P=\sum_n|\hat{\Psi}_{n,n}\rangle\langle\hat{\Psi}_{n,n}|$ as the projection over the kernel of
$L_H,$ $Q=\openone-P,$  and $\delta_t (x) :=[\sin(x t / 2)/(x/2)]^2,$ one finds that the operator metric  is given by the sum
of two distinct terns:
$\chi_\rho=\chi_\rho^{(1)}+\chi_\rho^{(2)},$ where
\begin{equation}
\chi^{(1)}_\rho=\| \delta_t (L_H)Q|H^\prime\rangle\|_\rho^2;\;\frac{\chi^{(2)}}{t^2}= \| P|H^\prime\rangle\|_\rho^2 -|\langle \openone, H^\prime\rangle_\rho|^2
\label{chirho}
\end{equation}

{\em Operator fidelity and random matrix theory}--
From the last Eqs. we see that, in general  $\chi_\rho$  depends on $t$, $H_{\lambda}$ and on
its derivative $H^\prime=\partial H_{\lambda}/\partial\lambda.$ We would like  to start our analysis of the chaos-related properties of $\chi_\rho$  by introducing a $t$-dependent quantity
$\hat{\chi}_\rho$  that, {\em contains information just on the Hamiltonian $H_{\lambda}$}. A possible way to achieve this goal is  i)  replace $H^\prime$ in the operator metric by a perturbation  $V$ that is assumed to be drawn by a Gaussian ensemble, with measure ${\cal D}[V],$ of {\em random matrices}\cite{RandomMtx}, ii) take $\hat{\chi}_\rho(t,H):=\int {\cal D} [V] \chi_\rho(t,H,V)$ as the average $\chi_\rho$ over the ensemble of $V$'s.

More precisely, we will assume that the the perturbation matrix elements $V_{n,m}$ are gaussian random variables
satisfying the relations
$\int {\cal D}[V] V_{n,m}\overline{V}_{p,q}\sim \delta_{n,p} \delta_{m,q}
$
The key step is the following fact:
If $R_\rho$ denotes the superoperator $X\rightarrow X\rho$ one has
$\int {\cal D}[V] |V\rangle\langle V|\sim R_\rho$ \cite{lemma} .
 Now, by using this result, and writing
$\hat{\chi}_\rho^{(1)}={\rm{Tr}}\left[ |V\rangle\langle V| \delta_t(L_H)Q\right]$  and
$\hat{\chi}_\rho^{(2)}={\rm{Tr}}\left[ |V\rangle\langle V| (P-|\openone\rangle\langle\openone|)\right],$
one finds
\begin{eqnarray}
\hat{\chi}^{(1)}_\rho(t,H) \sim {\rm Tr}\left [ R_\rho\delta_t(L_H)Q\right ],
\;
\hat{\chi}^{(2)}_\rho(t,H)\sim t^2(1-\rm{tr}\rho^2)
\nonumber
\end{eqnarray}
From these relations we clearly see that the two terms of the operator fidelity metric   behave quite differently upon averaging over the perturbation $V:$ on the one hand $ \hat{\chi}^{(2)}_\rho$ looses any direct connection with the Hamiltonian $H,$ just the purity of the state $\rho$ is returned, on the other hand the averaged $\hat{\chi}^{(1)}_\rho$ still shows an highly non-trivial dependence on the spectral features of $H$.
In the rest of the paper we are going to focus  on this latter term.

Let us make the content of the $\hat{\chi}^{(1)}_\rho$  explicit by resorting once again to the eigen-operator basis $|\hat\Psi_{n,n}\rangle$ of $L_H,$ from (\ref{chirho}) we find
$\hat{\chi}^{(1)}_\rho(t,H)\sim  \sum_{n\neq m}\rho_{n,n}\label{chi_mn}\delta_t(\Delta_{n,m})
$
$\Delta_{n,m}:=E_{n}-E_m$. This equation shows that the  perturbation averaged $\chi^{(1)}$ depends on the
the of level spacing distribution of $H.$ In particular -- in view of the property  $\lim_{t\to\infty} t^{-1} \delta_t(x)=2\pi\delta(x)$ --
for sufficiently large $t$ we expect contributions from  small $\Delta_{n,m}$ i.e., by  almost crossing levels, to dominate the operator metric.
 The key observation at this point is that one of the possible characterization of the presence of chaos in quantum system stems from the {\it level spacing analysis} \cite{Chaos,DickeChaos}. In this context, the properly normalized spectrum \cite{Unfold} is analyzed in terms of the spacing between consecutive energy levels $S_n=E_{n+1}-E_n.$
If  $P(S)$ denotes the probability that two nearest-neighbouring levels have an energy difference $S$  then the behaviour $P(S)$ for $S\rightarrow 0$ encodes the main features distinguishing a chaotic system from a regular one. Indeed, the distributions characterizing the regular spectra are Poissonian, allowing for a non vanishing probability to have $S=0$ spacing i.e., level crossings. In the chaotic case one has Wigner-like distributions $P_W(S)\propto S^{\nu} \exp{-S^2}$; their small spacing behaviour, in view of symmetry, is dominated by $S^\nu$, with $\nu >0.$ This entails for the phenomenon of {\em level repulsion} i.e., suppressed level crossings.

The above remarks together put us now in the position to formulate the main result of this paper in the form of a  Conjecture: {\em for sufficiently large $t$ the operator metric  $\chi^{(1)}_\rho$   is finite for regular quantum Hamiltonian and (almost) vanishing for quantum chaotic ones.}

To provide further support to this conjecture let us consider the {\em average} behavior of $\hat{\chi}^{(1)}_\rho$ with respect $H$
i.e., $\widetilde{\chi}^{(1)}_\rho(t):=\int {\cal D}[H]\hat{\chi}^{(1)}_\rho(t,H).$ Since $\hat{\chi}^{(1)}_\rho(t,H)$ is a spectral function integration  over the eigenvectors of $H$ can be readily performed  and one is left with \cite{RandomMtx}
\be
\widetilde{\chi}^{(1)}_\rho(t)\sim \int e^{ -\sum_n E_n^2} \prod_{n<m} dE_n dE_m (E_n-E_m)^\nu \hat{\chi}^{(1)}_\rho(t,H)\nonumber
\ee
 The exponent $\nu$ tells apart ensembles of regular, $\nu=0,$ from chaotic, $\nu>0$ Hamiltonians. From the last equation, and by resorting to  the explicit representation  $\hat{\chi}^{(1)}_\rho$ above we conclude
$\lim_{t\to\infty}   \widetilde{\chi}^{(1)}_\rho(t)/t\neq 0 $ for {regular} $H$
and
$\lim_{t\to\infty}   \widetilde{\chi}^{(1)}_\rho(t)/t\approx 0$ for {chaotic} $H.$
Assuming {\em typicality} i.e., typical and average behavior coincide, in $H$ these Eqs  are nothing but a formulation of the {\em Conjecture} above.

Of course none of the above arguments is rigorous, for as they depend on a set
of unproven assumptions e.g., RMT ensembles physical relevance, typicality, moreover they somehow neglect the potential
sensitivity on $\rho$ and finally rely on a "naive" large $t$ limit. Nevertheless  the conceptual content of the {\em Conjecture} is quite compelling and intuitive at the same time: typical regular Hamiltonians have a much higher susceptibility  against random perturbations than chaotic ones.
In other words, a random perturbation to a regular quantum evolution
results in new evolutions that has typically a much greater information-theoretic distance from the original one than if the same perturbation were applied to a quantum chaotic evolution.

{\em Operator fidelity and transition to chaos: the Dicke model}-- In this section we complement the above theoretical  analysis of the operator fidelity
 with a numerically study.  We will use the fidelity metric as a tool to study  the transition from regular to a chaotic regime in the Dicke model \cite{Dicke,DickeChaos}.
This model consists of a set of $N$ identical spin $1/2$ atoms with atomic level splitting $\omega_0$ placed in an ideal cavity that collectively interact with a single bosonic field described by the operators $\{a,a^\dagger\}=1$ and characterized by the frequency $\omega=\omega_0$ (resonance).
The  Dicke Hamiltonian reads
\begin{eqnarray}
H = \omega_0 J_z + \omega a^\dagger a
     + \frac{\lambda}{\sqrt{2j}} \left(a^\dagger + a \right )(J_+ + J_-).
\label{Eq.: DickeHam}
\end{eqnarray}
where $J_z=\sum_{i=1}^N \sigma_i^z$ and $J_{\pm}=J_x\pm J_{y}=\sum_{i=1}^N \sigma_i^{\pm}$ are collctive spin operators. The pseudo-spin length is fixed ($j=N/2$) and we have that the interaction is weighted by  $1/\sqrt{2j}=1/\sqrt{N}$.
The Dicke model is integrable at finite sizes only if, for small $\lambda$, one implements the rotating wave approximation (RWA) i.e., if one neglects the terms $a^\dagger J_+, a J_-$. In the thermodynamical limit (TDL) $N\rightarrow \infty $ \cite{Dicke, DickeChaos} it is integrable and it exhibits a quantum phase transition in correspondence of $\lambda=0.5$. This critical value separates the "normal" phase ($\lambda <0.5$) and the "super-radiant" phase and
the related  phase transition can be appropriately described in the context of the state fidelity approach \cite{paukozanardi}.

Here we are interested to the finite size instance of the model that, in absence of the RWA is not integrable and exhibits a regular to chaos cross-over. Indeed, the only (known) symmetry property is described by the parity operator $\bf{P}=\exp[i\pi \hat{N}]$, where $\hat{N}=a^\dagger a+ J_z +j$ is the operator that counts the number of total quanta present in the system. In order to characterize the quantum chaotic behaviour of (\ref{Eq.: DickeHam}) one can then resort to the study of the properties of the spectrum relative to the odd (even) subspaces. Indeed, in \cite{DickeChaos} the authors showed that in correspondence of the value of $\lambda=0.5$, which is critical in the TDL, one can observe a transition from a regular region ($\lambda<0.5$) characterized by Poisson-like level spacing distributions, to a chaotic region ($\lambda>0.5$) where the distributions are Wigner-like with $\nu=1$.

The actual system we used in our simulations  is a set of $N=20$ atoms coupled with a bosonic bath. To make computations feasible we have to introduce  a cut-off in the
--infinite dimensional -- bosonic state space. This cut-off has to be chosen in such a way that the bosonic system still operates as a bath for the atoms.
We thus choose to include the first $M=128$ bosonic energy levels; the total Hilbert space have thus dimension $d=4032$.

\begin{figure}
\includegraphics[height=3.5cm, width=8cm, viewport= 18 22 320 200,clip]{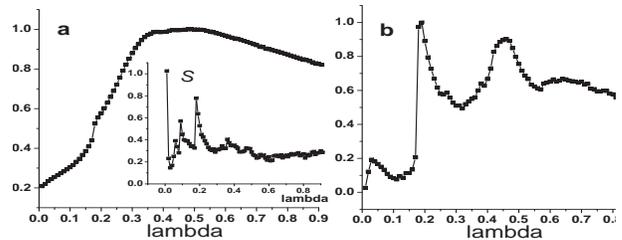}
\caption{Plots of $\chi^{(1)}_\rho(T,\lambda)$ corresponding to $T=1$ (a) and $T=10$ (b). The state is $\rho=\openone/d$. Inset: plot of the relative entropy $\mathcal{S}(P_\lambda||P_W)$ }
\label{Fig.: Suscept beta=0 T1T10}
\end{figure}

Let us first consider the level spacing distribution  $P_\lambda(S)$ of $H(\lambda)$. A  characterization of  $P_\lambda(S)$ can be given by its statistical distinguishability with respect to the Wigner distribution $P_W(S)$ ($\nu=1$); this statistical distance can be measured by the relative entropy $\mathcal{S}(P_\lambda||P_W):=\sum_n P_\lambda(S_n) \log[P_\lambda(S_n)/P_W(S_n)]$ \cite{RelEnt}. In the inset of Fig.
(\ref{Fig.: Suscept beta=0 T1T10}$.a$) we see that  in the chaotic chaotic region the relative entropy is uniform and small i.e., $P_\lambda(S)\approx P_W(S)$, whereas in the regular one
we have in general higher values and a more complex behavior since the degree of distinguishability varies with $\lambda$. The transition point being roughly at $\lambda=0.5.$
%

\begin{figure}
\includegraphics[height=3cm, width=8cm, viewport= 18 22 320 160,clip]{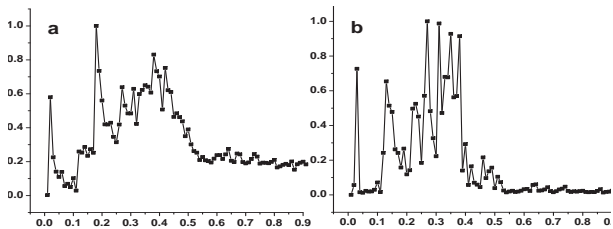}
\caption{Plots of $\chi^{(1)}_\rho(T,\lambda)$ corresponding to $T=100$ (a) and $T=1000$ (b). The state is $\rho=\openone/d$.}
\label{Fig.: Suscept beta=0 T100T1000}
\end{figure}

We now analyze the numerical results for
$\chi^{(1)}_\rho(t,\lambda)=  \sum_{n\neq m}\rho_{n,n}|\langle n|H^\prime|n\rangle|^2\delta_t(\Delta_{n,m})\nonumber$
where $H'=\partial_\lambda H={\sqrt{2j}} \left(a^\dagger + a \right )(J_+ + J_-)$ implements the perturbation to $H$.
We have first chosen $\rho=\openone/d$ i.e., scalar product in (\ref{Eq.: op fid}) is Hilbert-Schmidt, this corresponds to an infinite-temperature thermal state over the truncated working space. In Fig. (\ref{Fig.: Suscept beta=0 T1T10}) and (\ref{Fig.: Suscept beta=0 T100T1000}) we show $\chi^{(1)}_\rho(t=T,\lambda)$ (normalized to its maximal value $\max_{\lambda} \chi^{(1)}_\rho(T,\lambda)$) for different $T$'s.
These plots clearly show that for growing values of $T$ the behavior of $\chi^{(1)}_\rho(t,\lambda)$ changes significantly. For $T>1$ the response of the system to the infinitesimal change of parameter $\lambda$, as described by $\chi^{(1)}_\rho(t,\lambda)$, clearly exhibits two markedly different behaviour for $\lambda<0.5$ and $\lambda>0.5$. Indeed this behaviour is coherent with the conjecture described in the previous paragraphs: the long time response of the system is characterized by the sensitivity of  $\chi^{(1)}_\rho(t,\lambda)$ to the $S\rightarrow 0$ part of the spectrum. In particular, in the regular region, consistently with the relative entropy  analysis, the behaviour is non uniform with $\lambda$ and the (non uniform) presence of level crossings give rise to a pronounced response of $\chi^{(1)}_\rho$ characterized by a sequence of high peaks. In contrast, in the chaotic region the level repulsion phenomenon is reflected in a response which is uniform and low for $\lambda>0.5$

\begin{figure}[h!]
\includegraphics[height=3cm, width=8cm,viewport= 18 25 320 150,clip]{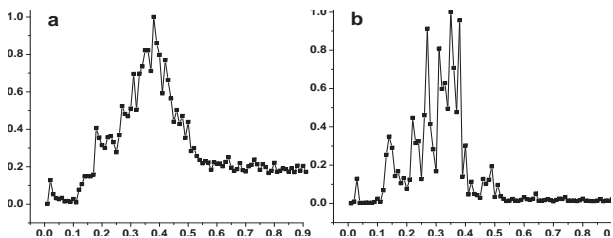}
\caption{Left (Right) plot of $\chi^{(1)}_\rho(t,\lambda)$ corresponding to $t=100 (t=1000)$. The state is $\rho=\exp{(-\beta H)}/{\rm Tr}[\exp{(-\beta H)}]$  where $\beta=0.014$.}
\label{Fig.: Suscept beta neq 0}
\end{figure}

We finally complete our analysis by showing in Fig. (\ref{Fig.: Suscept beta neq 0}) how $\chi^{(1)}_\rho(t,\lambda)$ behaves by choosing in (\ref{Eq.: op fid}) the thermal state $\rho=\exp{(-\beta H)}/{\rm Tr}[\exp{(-\beta H)}]$ with finite $\beta.$.
In principle the introduction of a finite temperature, by inducing a relative weight among the energy levels could spoil their chaotic vs regular distributions.
 Here we have chosen the inverse temperature $\beta=0.014$; this corresponds to fixing the ratio between the thermal probabilities relative to the ground and the highest energy state $\exp{[-\beta (E_{min}-E_{max})]}\approx 0.05$. Moreover we checked that for this $\beta$  the spectral features of $P(S)$ relevant to chaos are not washed out.
Fig. (\ref{Fig.: Suscept beta neq 0}) shows that  the $T=100$ and $T=1000$ behaviour of $\chi^{(1)}_\rho(t,\lambda)$ is again markedly different in the two regions; the response of the system is enhanced by the presence of level crossings in the regular region, while it is suppressed in the chaotic region.  These results are again consistent with our conjecture and suggest that the operator fidelity metric might well work as an order parameter for quantum chaos in appropriately chosen temperature intervals.

{\em Conclusions}-- In this paper we have  tackled  the problem of characterizing the chaotic properties of quantum systems by means of an information-geometrical tool: the operator fidelity metric. The results of our analysis are twofold. At a purely analytical level, by means of techniques drawn by random matrix theory, we have formulated and substantiated the conjecture that the operator fidelity metric may provide an order parameter  for  the regular to chaotic cross-over. The intuition being that a generic chaotic system,  at variance with a regular system,  is resilient with respect to random perturbations.
We have then shown how the operator fidelity approach can be  used as a tool to numerically identify the regular to chaotic cross-over in a relevant many-body system  i.e., the Dicke model.
We believe that the results obtained in this paper while preliminary, are promising and  give rise to  several interesting questions. Assessing the generality and efficiency of the methods we introduced and unveiling their relations with other approaches to quantum chaos e.g.,  \cite{casati08} is the subject of ongoing investigation.


\begin{thebibliography}{99}
\bibitem{woo-bra-ca} W. K. Wootters, Phys. Rev. D {\bf{23}}, 357 (1981).
S. L. Braunstein and C. M. Caves, Phys. Rev. Lett. {\bf{72}}, 3439 (1994).
\bibitem{paukozanardi} P.~Zanardi and N.~Paunkovic, Phys. Rev. E {\bf 74}, 031123 (2006);
P. Zanardi, P. Giorda and M. Cozzini, Phys. Rev. Lett. {\bf{99}}, 100603 (2007);   L. Campos Venuti, P. Zanardi, Phys. Rev. Lett. {\bf{99}}, 095701 (2007)
\bibitem{many-body} P. Zanardi, M. Cozzini and P. Giorda, J.~Stat.~Mech.~L02002 (2007):
M. Cozzini, P. Giorda and P. Zanardi, Phys.~Rev.~B \textbf{75}, 014439 (2007);
M.~Cozzini, R.~Ionicioiu and P.~Zanardi, Phys. Rev. B {\bf{76}}, 104420 (2007);
P. Buonsante and A. Vezzani, Phys.~Rev.~Lett.~\textbf{98}, 110601 (2007);
M.-F. Yang, arXiv:0707.4574;   H.-Q. Zhou, J.-H. Zhao and B.Li, arXiv:0704.2940; H.-Q. Zhou, arXiv:0704.2945;
Y.-C. Tzeng, M. -F. Yang, arXiv:0709.1518.
%
\bibitem{finite-T} P. Zanardi, H.-T Quan, X.-G. Wang, and C.-P. Sun, Phys. Rev. A {\bf{75}}, 032109 (2007);
P. Zanardi, L. Campos Venuti, P. Giorda, Phys. Rev. A {\bf{76}}, 062318 (2007); D. F. Abasto, N. T. Jacobson, P. Zanardi,  arXiv:0711.139
\bibitem{opfid} X. Wang, Z. Sun, Z. D. Wang,
arXiv:0803.2940; X. Lu, Z. Sun , X. Wang , and P. Zanardi,Phys. Rev. A {\bf{78}}, 032309 (2008)
\bibitem{distin}
If  $\rho=\sum_i p_i \ketbra{i}{i}$ one considers the state $\ket{\Psi_\rho} =\sum_i \sqrt{p_i} \ket{i}\otimes\ket{i} \in {\cal H}\otimes {\cal H}$ then , for each $A \in {\cal L}({\cal H})$ one  defines $\ket{A}\doteq (A\otimes \openone)\ket{\Psi_\rho}.$ The operator fidelity is then the (bi-partite) state fidelity:  ${\cal F}_\rho(X,Y)=\langle X|Y\rangle.$
\cite{opfid}.

\bibitem{Chaos} F. Haake, {\it Quantum Signatures of Chaos}   (Springer 2001); H-J. Stockmann, {\it Quantum Chaos: an introduction} (Cambridge University Presse 1999).
\bibitem{RandomMtx} M. L. Mehta, {\it Random matrices}, (Elsevier 2004)
\bibitem{QCh-rev}
D.L. Shepelyansky, Physica D {\bf{8}}, 208 (1983); G. Casati
et al., Phys. Rev. Lett. {\bf{56}}, 2437(1986).
\bibitem{QCh-stab}A. Peres, Phys. Rev. A {\bf{30}}, 1610(1984)
\bibitem{casati08} G. Benenti, G Casati,  arXiv:0808.3243
 \bibitem{Dicke}  R. H. Dicke, Phys. Rev. {\bf 93}, 99 (1954);
 \bibitem{DickeChaos} C. Emary and T. Brandes, Phys. Rev. Lett. 90, 044101 (2003);C. Emary and T. Brandes, Phys. Rev. E 67, 066203 (2003)
\bibitem{losch} H.T. Quan, Z. Song, X.F. Liu, P. Zanardi, C.P. Sun, Phys. Rev. Lett. {\bf{96}}, 140604 (2006)
\bibitem{Unfold} T. Guhr, A. M¨uller-Groeling, and H. A. Weidenm¨uller, Phys. Rep. 299, 189 (1998).
\bibitem{RelEnt} T. M. Cover and J. A. Thomas, Elements of Information Theory (Wiley, New York, 1991).
\bibitem{kur} J. Kurchan,  cond-mat/0007360
\bibitem{lemma} To prove this result is sufficient to take the matrix elements of its both sides of with respect the eigen-operator basis
 $|\hat\Psi_{n,m}\rangle$ and perform the gaussian averages

\end{thebibliography}
\end{document}